\documentclass{aa}
\usepackage{graphicx}

\begin{document} 


   \title{Cu and Zn in the early Galaxy
          }

   \titlerunning{Cu and Zn in the early Galaxy}

   \author{G. Bihain\inst{1}
      \and G. Israelian\inst{1}
      \and R. Rebolo\inst{1}\fnmsep\inst{2}
      \and P. Bonifacio\inst{3}
      \and P. Molaro\inst{3}}
   \authorrunning{G. Bihain et al.}
   \institute{Instituto de Astrof\'{\i}sica de Canarias, c/ V\'{\i}a
              L\'actea,
              s/n, 38205  La Laguna, Tenerife, Spain\\
              \email{gbihain@ll.iac.es, gil@ll.iac.es and rrl@ll.iac.es}
              \and
	      Consejo Superior de Investigaciones Cient\'{\i}ficas, Spain
	      \and
	      Istituto Nazionale di Astrofisica, Osservatorio Astronomico di Trieste, via G. B. Tiepolo 
	      11, 34131 Trieste, Italy\\
              \email{molaro@ts.astro.it and bonifacio@ts.astro.it}}
   \offprints{G. Bihain, \email{gbihain@ll.iac.es}}
   
   \date{Received date; accepted date} 

   \abstract{We present Cu and Zn abundances for 38 FGK stars,
mostly dwarfs, spanning a metallicity range between solar and
[Fe/H] =~$-$3. The abundances were obtained using Kurucz's local
thermal equilibrium (LTE) model atmospheres and the near-UV
lines of \ion{Cu}{i} 3273.95~{\AA} and \ion{Zn}{i} 3302.58~{\AA}
observed at high spectral resolution. The trend of [Cu/Fe]
versus [Fe/H] is almost solar for [Fe/H] $>$~$-$1 and then
decreases to a plateau $\langle$[Cu/Fe]$\rangle$ =~$-$0.98 at
[Fe/H] $<$~$-$2.5, whereas the [Zn/Fe] trend is essentially
solar for [Fe/H] $>$~$-$2 and then slightly increases at lower
metallicities to an average value of $\langle$[Zn/Fe]$\rangle$
=~+0.18. We compare our results with previous work on these
elements, and briefly discuss them in terms of nucleosynthesis
processes. Predictions of halo chemical evolution fairly
reproduce the trends, especially the [Cu/Fe] plateau at very low
metallicities, but to a lower extent the greater [Zn/Fe] ratios
at low metallicities, indicating eventually missing yields.

	 }

  \maketitle

\section{Introduction}

Massive, low- and intermediate-mass stars, type II and type Ia
supernovae are likely contributors to the Galactic chemical
evolution of Cu and Zn. However, the nucleosynthesis mechanisms
and the relative contributions of the nucleosynthesis sites for
these near- iron-peak nuclei are still uncertain. The abundances
of Cu in metal-poor stars with [Fe/H] $>$~$-$1.7 have been
comprehensively reviewed by Peterson (\cite{pet81}), with no
significant deviation from the solar ratio [Cu/Fe] =~0. Further
studies by Luck \& Bond (\cite{luc85}),  Sneden \& Crocker
(\cite{sne88}) and Sneden et al. (\cite{sne91}) show that, while
the [Zn/Fe] ratio remains essentially solar, there is in fact a
clear decrease in [Cu/Fe] at lower metallicities, consistent
with previous results for the metal-poor globular clusters M13
and M92 by Cohen (\cite{coh78}, \cite{coh79}, \cite{coh80}).
This departure from the solar ratio is as particular as for the
iron peak elements Cr, Mn and Co; [Co/Fe], however, is
increasing at lower metallicities (e.g. Cayrel et al.
\cite{cay03}). More recent studies (Mishenina et al.
\cite{mis02}; Simmerer et al. \cite{sim03}) show that the trend
of Cu is non-linear in metal-poor stars and globular clusters,
what presumably results from the superposition of Fe-independent
(``primary'') and Fe-dependent (``secondary'') nucleosynthesis
mechanisms. While a general consensus is set on the early
contribution by short-lived massive stars, there are many
uncertainties concerning how these stars produce Cu and Zn, and
how their yields lead to the observed low-metallicity tails of
the trends.\\

The  above-mentioned studies are performed mostly with optical
lines\footnote{In a study of stellar abundances in the thick
disc of the Galaxy (Prochaska et al. \cite{pro00}), the IR line
\ion{Cu}{i} 8092~{\AA} is also used.} (\ion{Cu}{i} at
5105.54~{\AA}, 5218.20~{\AA} and 5782.12~{\AA}; \ion{Zn}{i} at
4722.16~{\AA}, 4810.53~{\AA} and 6362.35~{\AA}). These lines
become very weak in the spectra of very metal-poor stars.
Therefore we decided to explore the use of the UV lines 
\ion{Cu}{i} 3273.95~{\AA} and \ion{Zn}{i} 3302.58~{\AA}, which
are quite intense in the solar spectrum and, in the case of the
Cu line, potentially detectable at very low metallicities
([Fe/H] $<$~$-$3). In the following sections, we will discuss 
the observations, their analysis, and how these lines are indeed
useful for investigating the chemical evolution of copper and
zinc at low metallicities.\\

\section{Observations}

\begin{figure*}
\centering
  \includegraphics[width=17cm]{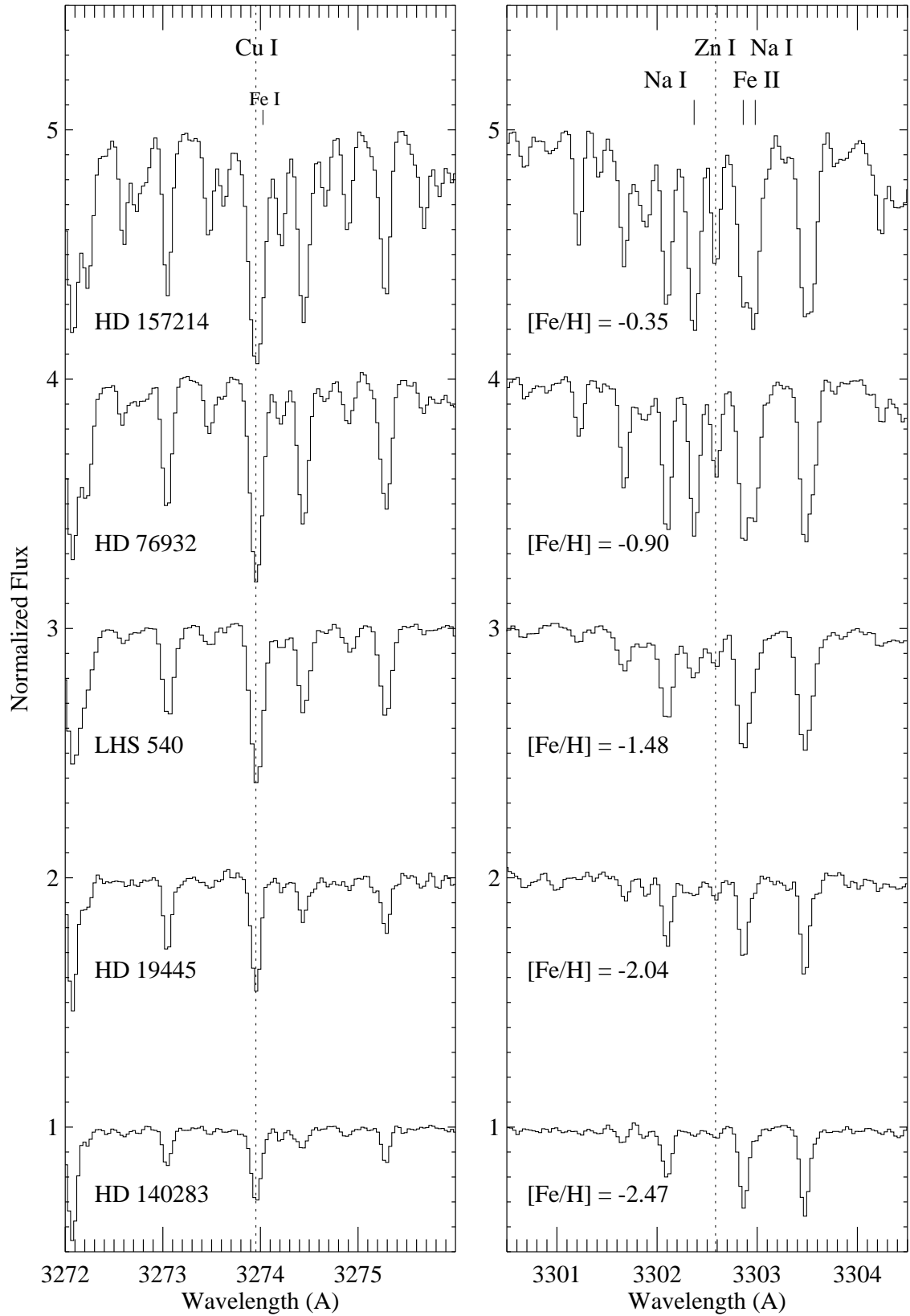}
  \caption{Spectra of UV copper and zinc lines in metal-poor dwarfs.}
  \label{obs_spectra}
\end{figure*}

 The studied sample contains 38 FGK metal-poor stars,
principally dwarfs. Most of the high-resolution spectra were
obtained at the 4.2~m WHT/UES (La Palma) during 1992--1993.
Spectra of the stars \object{HD 3795}, \object{LHS 540},
\object{HD 211998}, \object{HD 166913}, \object{HD 218502} and
\object{HD 128279} were obtained at the 3.6~m/CASPEC (La Silla)
during 1993; and of the stars \object{HD 22879} and \object{HD
132475} at the 3.9~m AAT/UCLES (Siding Spring) during
1990--1991. The spectra of the very metal-poor stars \object{LP
815-43}, \object{G275-4} and \object{G64-12} were obtained at
the 8.2~m VLT Kueyen/UVES (Paranal) during 1999--2000 (see
details in Israelian et al. \cite{isr01}). Finally, the spectra
of the very metal-poor giants \object{HD 2796} and \object{BD
$-$18$\degr$5550} were obtained at the WHT/UES during 2000, as
described in Israelian \& Rebolo (\cite{isr01}). The \'echelle
spectra covered the spectral range 3025--3825~{\AA}, with
spectral resolution $R \sim$~40\,000--60\,000 and
signal-to-noise ratio $S/N \sim$~80--150 at 3300~{\AA}. All the
data were reduced using standard IRAF\footnote{IRAF is
distributed by the National Optical Astronomical Observatories,
which is operated by the Association of Universities for
Research in Astronomy, Inc., under contract with the National
Science Foundation.} routines for bias, scattered light, flat
field correction and wavelength calibration. We present in
Fig.~\ref{obs_spectra} the spectra of five stars with
different metallicities. The normalization of the spectra was
achieved by fitting a low-order polynomial to regions of
continuum previously identified in the solar spectrum (Kurucz et
al. \cite{kur84}), in the neighborhood of the Cu and Zn lines.

\section{Spectral synthesis and stellar parameters}

\begin{figure*}
\centering
\includegraphics[width=17cm,height=10cm]{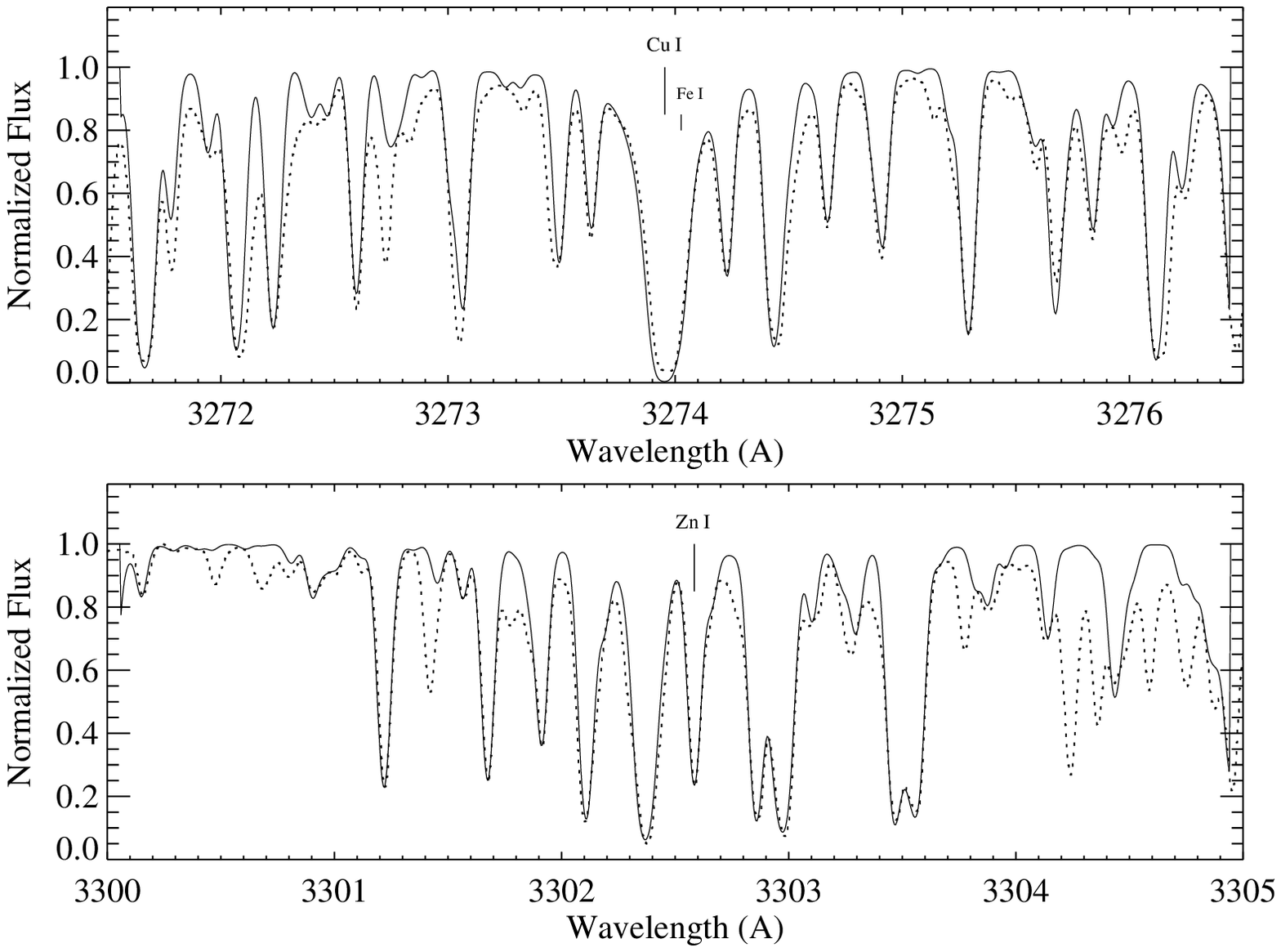}
\caption{Expanded plots of the solar spectral regions ({\sl
dotted lines}) fitted with the synthetic spectrum ({\sl solid
lines}).}
\label{sunfit}
\end{figure*}

Model atmospheres with ``overshooting'' were chosen from the
ATLAS9 grid (Kurucz \cite{kur93}) by standard interpolation
through [Fe/H], $T_{\rm eff}$ and log $g$. In order to analyze
the blended lines of copper and zinc, 5~{\AA} wide synthetic
spectra were computed with the LTE code MOOG (Sneden
\cite{sne73}). The atomic parameters of the \ion{Cu}{i}
resonance lines at 3247.54~{\AA} (log~$gf$ =~$-$0.062, $\chi$
=~0.000) and at 3273.95~{\AA} (log~$gf$ =~$-$0.359, $\chi$
=~0.000), and those of the \ion{Zn}{i} line at 3302.58~{\AA}
(log~$gf$ =~$-$0.057, $\chi$ =~4.030) were taken from  VALD-2
(Kupka et al. \cite{kup99}).  The wavelengths and the relative
strengths of the hyperfine components of the copper isotopes
\element[][63]{Cu} and \element[][65]{Cu} were obtained from
Kurucz (\cite{kur03}, private communication), and the solar
isotopic ratio was adopted from Anders \& Grevesse
(\cite{and89}) for all the stars. The oscillator strengths of
the lines were calibrated with the solar spectrum, using a model
atmosphere with overshooting from Kurucz (\cite{kur92}, private
communication) for the parameters $T_{\rm eff,\sun}$ =~5777~K,
log~$g_{\sun}$ =~4.438, $V_{\rm t,\sun}$ =~1,0~km~s$^{-1}$ and
the Unsold damping approximation without enhancement. The
calibration provided log~$gf$ $=$~$-$0.167 for the \ion{Zn}{i}
line (see Fig.~\ref{sunfit}). In the case of the \ion{Cu}{i}
lines log~$gf$ was not modified, because its modification
strongly affected the suitable fit to the wings (as obtained
from the {\sl cog} driver of MOOG, the lines were in the linear
damping part of the curve of growth) and did not improve the fit
to the cores. In fact, the cores of such lines are usually
formed close to the chromospheric temperature minimum, which
exceeds that of the upper photosphere. Since no chromosphere is
included in our model, these cores could not be reproduced like
in the observed spectrum.\\

Stellar parameters for the model atmospheres were obtained as
follows. The effective temperatures ($T_{\rm eff}$) were
estimated using the Alonso et al. (\cite{alo96}, \cite{alo99b})
calibrations versus $V-K$ colours for dwarfs and giants,
respectively. The $V$ magnitudes were taken from the Hipparcos
Catalog (Perryman et al. \cite{per97}), Alonso et al.
(\cite{alo94}) or else from the SIMBAD database. The $K$
magnitudes were taken from Alonso et al. (\cite{alo94},
\cite{alo98}), or else from Carney \& Aaronson (\cite{car79};
\object{HD 170153}, \object{HD 165908}, \object{HD 225239} and
\object{HD 157214}) and Laird et al. (\cite{lai88};
\object{G170-47}). For \object{HD 166913}, a $K_{\rm s}$
magnitude was taken from the 2MASS Catalogue, and then converted
to $K$ using a relation determined from a comparison of the
magnitudes in these bands for similar stars in the sample. Since
no accurate $K$ or $K_{\rm s}$ values were available for the
stars \object{HD 211998} and \object{HD 3795}, the $T_{\rm
eff}$--($B-V$) calibrations from Alonso et al. (\cite{alo96},
\cite{alo99b}) were used, with ($B-V$) from the Hipparcos
Catalogue or the SIMBAD database. LTE metallicities were adopted
from the literature, mainly from Fulbright (\cite{ful00}),
Mishenina et al. (\cite{mis02}) and Th\'evenin (\cite{the98}).
The differences in the values provided by these three sources
were typically not greater than 0.12~dex for the stars in
common, therefore justifying the use of an average metallicity
(see Table~\ref{atmparam}). Finally, the initial surface
gravities were taken from the same references, with the
exception of Th\'evenin \& Idiart (\cite{the99}) which was used
instead of Th\'evenin (\cite{the98}), because a higher priority
was given to their NLTE log~$g$ values (see comments below).
Stars with surface gravities log~$g$ $>$~3.5 were classified as
dwarfs, while stars with log~$g$ $<$ 2.5 were classified as
giants, so that the corresponding $T_{\rm eff}$ colour
calibration was used for each group. When stars had an
intermediate log~$g$ value, $T_{\rm eff}$ was adopted as the
average of the temperatures obtained with the calibrations for
dwarfs and for giants, respectively.\\

\begin{figure}
\resizebox{\hsize}{!}{\includegraphics{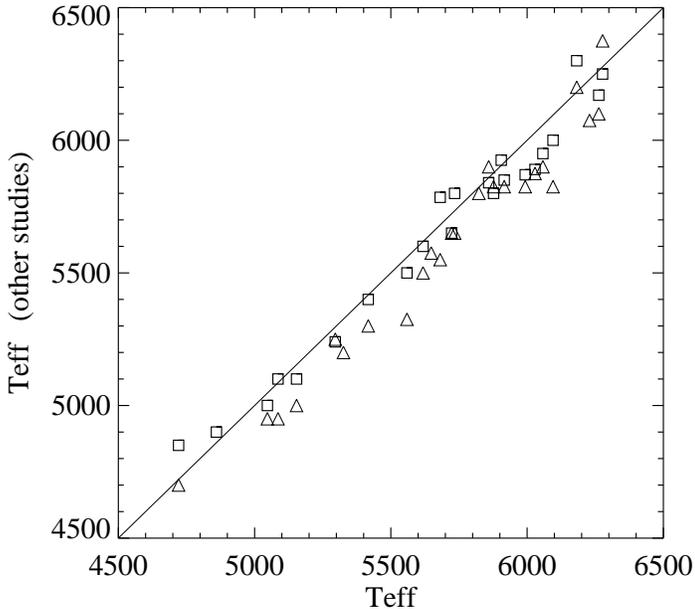}}
\caption{Comparison of the $T_{\rm eff}$ obtained in this study
with spectroscopic  $T_{\rm eff}$ from Fulbright ({\it
triangles}) and Mishenina \& Kovtyukh ({\it squares}).}
\label{Teff_tw-Mish2001} \end{figure}

 We found a very good agreement between our $T_{\rm eff}$ values
and the spectroscopic $T_{\rm eff}$ values from Mishenina \&
Kovtyukh (\cite{mis01}) for the 23 stars in common
($\langle$$T_{\rm eff}$$-$T$_{\rm eff, MK}$$\rangle$ =~+26~K,
with a standard deviation $\sigma$ =~75~K, see
Fig.~\ref{Teff_tw-Mish2001}). A comparison with the
spectroscopic $T_{\rm eff}$ values from Fulbright
(\cite{ful00}), however, showed  a systematic difference of
$\langle$$T_{\rm eff}$$-$T$_{\rm eff, F}$$\rangle$ =~+99~K with
$\sigma$ =~82~K (25 stars). Since the 21 stars for this latter
comparison were also in that with Mishenina \& Kovtyukh's
sample, this difference should be attributed to the different
methods used. Fulbright determines for the dwarfs an excitation
temperature (requiring the same iron abundance for Fe I lines
with high and low excitation potential), whereas Mishenina \&
Kovtyukh determine a temperature from the fits to the wings of
the H$_{\alpha}$ lines. Comparing with Th\'evenin's
(\cite{the98}) $T_{\rm eff}$ values, we found $\langle$$T_{\rm
eff}$$-$T$_{\rm eff,T}$$\rangle$ =~+79~K with $\sigma$ =~97~K
(23 stars). These are selected from the Cayrel de Strobel et
al.'s (\cite{cay92}) catalogue, and redetermined from colours in
the rare case that they are initially determined from Fe I lines
(Th\'evenin \cite{the03}, private communication). Comparing for
the remaining stars, we found similar values to those in the
references used for the metallicities, with a difference of less
than 80~K, apart for the star \object{HD 128279}, for which we
obtained 5130~K, and Gratton et al. (\cite{gra00}), 5394~K. This
latter value is derived from dereddened ($B-V$)$_{0}$ and
($b-v$)$_{0}$ colours using colour--$T_{\rm eff}$
transformations (Kurucz \cite{kur95}), and is higher, as for
nearly all the common stars with the present sample.\\

The surface gravities (log~$g$) were estimated using Hipparcos
parallaxes (Perryman et al. \cite{per97}) through the  relation
proposed by Nissen et al. (\cite{nis97}), [$g$] =~[$M$] +~4
[$T_{\rm eff}$] +~0.4~($M_{\rm bol}$ -~$M_{\rm bol,\sun}$),
where [$x$] =~log $x/{x_{\sun}}$. The solar bolometric magnitude
$M_{\rm bol,\sun}$ =~4.75~mag was taken from Allen
(\cite{all76}), and the bolometric magnitudes of the stars were
determined using interpolated bolometric corrections from the
grids given by Alonso et al. (\cite{alo99a}): $M_{\rm bol}$
=~$K$ +~BC($K$) +~5~log~$\pi$ +~5, where $\pi$ is the
trigonometric parallax. For the two stars without accurate $K$
and $K_{\rm s}$ magnitudes BC($V$) was interpolated from the
grids of Alonso et al. (\cite{alo95}, \cite{alo99b}). The
relative standard error in parallax, $\sigma(\pi)/\pi$, was
typically of 0.10 and smaller than 0.30. The stellar masses
($M$) were derived from their position in the (log $T_{\rm
eff}$, log $L/L_{\sun}$) plane, using the isochrones from
Bergbusch \& VandenBerg (\cite{ber01}). Most of the stars were
found to be in the main sequence and the subgiant domain, while
the remainder were in the red giant branch. The average stellar
mass for the sample was 0.78~$M_{\sun}$. The error in log $g$
arose mainly from the parallax uncertainty: an error of
$\sigma(\pi)/\pi$ =~0.10 implied (via the bolometric magnitude)
$\Delta$~log~$g$ =~0.10~dex, while typical errors in the mass
$\sigma(M)$ =~0.045~$M_{\sun}$ and in the effective temperature
$\sigma$($T_{\rm eff}$) =~75~K implied  errors of  only $\Delta$
log~$g$ =~0.025~dex.\\

\begin{figure}
\resizebox{\hsize}{!}{\includegraphics{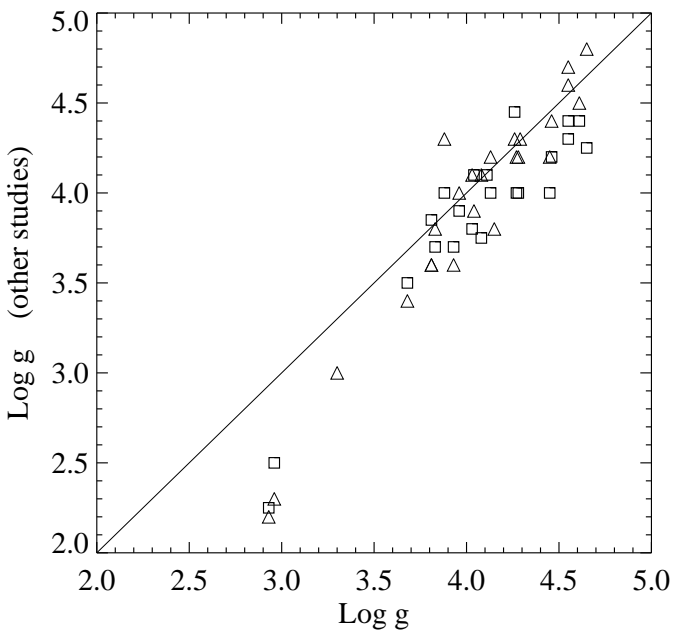}}
\caption{Comparison of the log~$g$ obtained in this study with
spectroscopic log~$g$ from Fulbright ({\sl triangles}) and
Mishenina \& Kovtyukh ({\sl squares}).}
\label{logg_tw-Fulb2000-Mish2001} \end{figure}

 The surface gravities of the dwarf stars agreed in general with
the LTE spectroscopic values from Fulbright (\cite{ful00}) and
Mishenina \& Kovtyukh (\cite{mis01}), see
Fig.~\ref{logg_tw-Fulb2000-Mish2001}. However, those of the very
metal-poor subgiants \object{HD 87140} and \object{G170-47} were
relatively different, with $\Delta$log~$g$ $>$~+0.40~dex. This
discrepancy between trigonometric and LTE spectroscopic surface
gravities in late-type stars is discussed by Nissen et al.
(\cite{nis97}) and Allende Prieto et al. (\cite{allp99}).
Moreover, trigonometric surface gravities agree better with NLTE
spectroscopic determinations (Th\'evenin \& Idiart
\cite{the99}). Since no NLTE value is published for \object{HD
87140}, we could only compare with a trigonometric one, for
instance that given by Gratton et al. (\cite{gra00}) and we
found a difference of 0.05~dex. For \object{G170-47}, we could
compare with the NLTE ``best choice'' determination from
Israelian et al. (\cite{isr01}) and found a difference of
0.07~dex. As for the remaining stars, we found  log~$g$ values 
very similar to those in the references we used for
metallicities, with differences of less  than 0.17~dex.\\

For the three very metal-poor dwarfs \object{G64-12},
\object{G275-4} and \object{LP815-43} observed with VLT/UVES, the
atmospheric parameters were taken from Israelian et al.
(\cite{isr01}). For the two very metal-poor giants \object{BD
$-$18$\degr$5550} and \object{HD 2796}, the spectroscopic log~$g$
and $T_{\rm eff}$ were taken from Israelian \& Rebolo
(\cite{isr01}). Finally, the microturbulence was fixed at $V_{\rm
t}$ =~1.0~km~s$^{-1}$ for the dwarfs and 1.5~km~s$^{-1}$ for the
giants.\\

\begin{table*} \caption[]{Stellar atmospheric parameters and Cu
and Zn abundances. Some parameters are from 1) Castro et al.
(\cite{cas99}) 2) Chen et al. (\cite{che00}) 3) Fulbright
(\cite{ful00}) 4) Spite \& Spite (\cite{spi82}) 5) Gratton et
al. (\cite{gra00}) 6) Mishenina et al. (\cite{mis02}) 7)
Th\'evenin (\cite{the98}) 8) Israelian et al. (\cite{isr01}) 9)
Israelian \& Rebolo (\cite{isr01}). Solar Cu and Zn abundances
are from Anders \& Grevesse (\cite{and89}).}

      \label{atmparam}
  $$
     {\small	
     \begin{tabular}{l l l l l l l l l l}
     	 \hline
	 Star &
	 $T_{\rm eff}$ &
	 log $g$ &
	 $V_{\rm t}$&
	 [Fe/H] &
	 Ref. &
	 [Cu/H] &
	 $\sigma$ &
	 [Zn/H] &
	 $\sigma$
	 \\
	& $_\mathrm{K}$ &  & $_\mathrm{km\,s^{-1}}$ & &&&&\\
	 \hline
	 \object{G64-12}             & 6318 (150)$^{8}$ &  4.20 (.3)$^{8}$ & 1.0  & -3.05$^{8}$     &c  & -4.00    & 0.18 & $<$-2.35 & --   \\
	 \object{BD -18$\degr$5550}  & 4668 (63)$^{9}$  &  1.5 (.3)$^{9}$  & 1.5  & -3.01$^{6}$     &b  & -3.94    & 0.20 & -2.81    & 0.17 \\
	 \object{G275-4}             & 6212 (150)$^{8}$ &  4.13 (.3)$^{8}$ & 1.0  & -2.99$^{8}$     &c  & -3.85    & 0.18 & $<$-2.34 & --   \\
	 \object{BD +3$\degr$740}    & 6229 (67)        &  4.15 (.27)      & 1.0  & -2.82$^{3,7}$   &b  & $<$-3.77 & --   & $<$-2.02 & --   \\
	 \object{LP 815-43}          & 6265 (125)$^{8}$ &  4.54 (.3)$^{8}$ & 1.0  & -2.74$^{8}$     &c  & -3.79    & 0.15 & $<$-2.29 & --   \\
	 \object{G170-47}            & 5154 (58)        &  2.93 (.28)      & 1.25 & -2.66$^{3,6}$   &b  & -3.76    & 0.19 & $<$-2.21 & --   \\
	 \object{HD 140283}          & 5723 (67)        &  3.68 (.06)      & 1.0  & -2.47$^{3,6,7}$ &b  & -3.42    & 0.14 & -2.27    & 0.14 \\
	 \object{BD +26$\degr$3578}  & 6263 (76)        &  3.93 (.21)      & 1.0  & -2.36$^{3,6,7}$ &b  & -3.01    & 0.14 & $<$-1.96 & --   \\
	 \object{HD 2796}            & 4860 (46)        &  1.80 (.2)$^{9}$ & 1.5  & -2.18$^{6,7}$   &b  & --       & --   & -2.03    & 0.17 \\
	 \object{BD +37$\degr$1458}  & 5326 (54)        &  3.30 (.23)      & 1.25 & -2.17$^{3}$     &b  & -2.87    & 0.14 & -1.82    & 0.12 \\
	 \object{HD 128279}          & 5130 (66)        &  2.85 (.22)      & 1.25 & -2.11$^{5}$     &a  & -3.11    & 0.20 & -2.31    & 0.17 \\
	 \object{HD 84937}           & 6277 (75)        &  4.03 (.09)      & 1.0  & -2.06$^{3,6,7}$ &b  & -2.96    & 0.14 & $<$-1.96 & --   \\
	 \object{HD 19445}           & 6095 (69)        &  4.45 (.05)      & 1.0  & -2.04$^{3,6,7}$ &b  & -2.69    & 0.14 & -1.69    & 0.12 \\
	 \object{HD 87140}           & 5086 (43)        &  2.96 (.31)      & 1.25 & -1.83$^{3,6}$   &b  & -2.28    & 0.21 & $<$-1.13 & --   \\
	 \object{HD 25329}           & 4721 (65)        &  4.65 (.06)      & 1.0  & -1.78$^{3,6,7}$ &b  & -2.23    & 0.20 & --       & --   \\
	 \object{HD 218502}          & 6182 (77)        &  4.08 (.08)      & 1.0  & -1.76$^{3,6,7}$ &a  & -2.51    & 0.17 & -1.61    & 0.12 \\
	 \object{HD 64090}           & 5417 (65)        &  4.55 (.06)      & 1.0  & -1.74$^{3,6,7}$ &b  & -2.19    & 0.20 & -1.64    & 0.16 \\
	 \object{HD 166913}          & 6181 (74)        &  4.18 (.07)      & 1.0  & -1.68$^{7}$     &a,d& -1.88    & 0.41 & -1.23    & 0.25 \\
	 \object{HD 132475}          & 5648 (69)        &  3.81 (.10)      & 1.0  & -1.60$^{3,7}$   &d  & -2.10    & 0.25 & -1.10    & 0.18 \\
	 \object{HD 188510}          & 5559 (65)        &  4.55 (.06)      & 1.0  & -1.57$^{3,6}$   &b  & -1.88    & 0.24 & -1.52    & 0.20 \\
	 \object{BD +23$\degr$3912}  & 5734 (67)        &  3.83 (.13)      & 1.0  & -1.50$^{3,6,7}$ &b  & -1.90    & 0.25 & --       & --   \\
	 \object{HD 94028}           & 6058 (72)        &  4.27 (.06)      & 1.0  & -1.49$^{3,6,7}$ &b  & -1.79    & 0.29 & -1.39    & 0.17 \\
	 \object{HD 211998}          & 5282 (180)       &  3.28 (.28)      & 1.25 & -1.48$^{7}$     &a,d& -2.10    & 0.34 & -1.43    & 0.14 \\
	 \object{LHS 540}            & 5993 (71)        &  3.88 (.20)      & 1.0  & -1.48$^{3,6}$   &a  & -1.63    & 0.26 & -1.33    & 0.12 \\
	 \object{HD 103095}          & 5047 (68)        &  4.61 (.05)      & 1.0  & -1.43$^{3,6,7}$ &b  & -1.83    & 0.16 & -1.38    & 0.11 \\
	 \object{HD 194598}          & 6029 (77)        &  4.28 (.07)      & 1.0  & -1.20$^{3,6,7}$ &b  & -1.45    & 0.32 & --       & --   \\
	 \object{HD 189558}          & 5681 (70)        &  3.81 (.08)      & 1.0  & -1.12$^{3,6,7}$ &b  & -1.42    & 0.21 & -1.02    & 0.16 \\
	 \object{HD 201891}          & 5916 (77)        &  4.26 (.05)      & 1.0  & -1.05$^{3,6,7}$ &b  & -0.95    & 0.26 & -0.85    & 0.12 \\
	 \object{HD 201889}          & 5618 (67)        &  4.04 (.08)      & 1.0  & -0.95$^{3,6,7}$ &b  & -1.00    & 0.21 & -0.80    & 0.16 \\
	 \object{HD 22879}           & 5823 (70)        &  4.29 (.04)      & 1.0  & -0.91$^{3}    $ &d  & -1.16    & 0.25 & -0.76    & 0.18 \\
	 \object{HD 76932}           & 5859 (72)        &  4.13 (.04)      & 1.0  & -0.90$^{3,6,7}$ &b  & -1.02    & 0.21 & -0.74    & 0.16 \\
	 \object{HD 6582}            & 5296 (67)        &  4.46 (.05)      & 1.0  & -0.86$^{3,6,7}$ &b  & -0.96    & 0.17 & -0.91    & 0.11 \\
	 \object{HD 134169}          & 5877 (72)        &  3.96 (.07)      & 1.0  & -0.81$^{3,6,7}$ &b  & -0.66    & 0.21 & -0.81    & 0.16 \\
	 \object{HD 3795}            & 5226 (185)       &  3.78 (.07)      & 1.0  & -0.70$^{1}$     &a  & -0.85    & 0.34 & -0.75    & 0.12 \\
	 \object{HD 165908}          & 5905 (42)        &  4.11 (.03)      & 1.0  & -0.67$^{6}$     &b  & -0.67    & 0.24 & -0.67    & 0.12 \\
	 \object{HD 170153}          & 5954 (42)        &  4.13 (.03)      & 1.0  & -0.65$^{2}$     &b  & -0.70    & 0.24 & -0.60    & 0.12 \\
	 \object{HD 225239}          & 5528 (44)        &  3.74 (.15)      & 1.0  & -0.5$^{4}$      &b  & -0.56    & 0.18 & -0.54    & 0.16 \\
	 \object{HD 157214}          & 5715 (43)        &  4.21 (.03)      & 1.0  & -0.35$^{7}$     &b  & -0.40    & 0.16 & -0.35    & 0.16 \\
	\hline \\
      \end{tabular}
      }
  $$
  a. CASPEC- 3.6 m Telescope b. UES- WHT c. UVES- VLT d. UCLES- AAT
\end{table*}

\subsection{Cu and Zn Abundances}
\label{Cu and Zn Abundances}

The abundances of Cu and Zn are listed in Table~\ref{atmparam}.
The wings of the intense \ion{Cu}{i} lines were fitted with the
synthetic spectrum by differential analysis with respect to the
\object{Sun}. Because these lines were in the damping or
saturation parts of the curve of growth for the stars with solar
temperature and metallicities $-$2 $<$~[Fe/H] $<$~0, the effects
of changing the abundances were carefully checked. In general an
uncertainty of 0.1~dex was adopted for the derived abundances.
For metallicities lower than $-$2 the lines were unsaturated,
and the adjustments could be achieved more precisely.\\

The uncertainties in $T_{\rm eff}$ and log~$g$ led to average
errors of respectively 0.11 and 0.02~dex in the Cu abundance,
and 0.035 and 0.02~dex in the Zn abundance, while the $V_{\rm
t}$ uncertainty led to an average error of 0.07 in the Cu
abundance and 0.05~dex in the Zn abundance. The main
contribution to the error came from the uncertainty in the
localization of the continuum, especially in the spectra of the
more metal-rich stars and in the noisy spectra; this uncertainty
implied an error of 0.1--0.2~dex in the abundances. Additionally
the noise in the copper and zinc lines led to an error of
0.05--0.1~dex. In the case of the \ion{Cu}{i}
$\lambda$3273.95~{\AA} line, an extra error of  0.01--0.04~dex
was taken into account for the uncertainty in the contribution
of the weak \ion{Fe}{i} $\lambda$3273.95~{\AA} line.\\

\begin{figure*} \centering
\includegraphics[width=17cm]{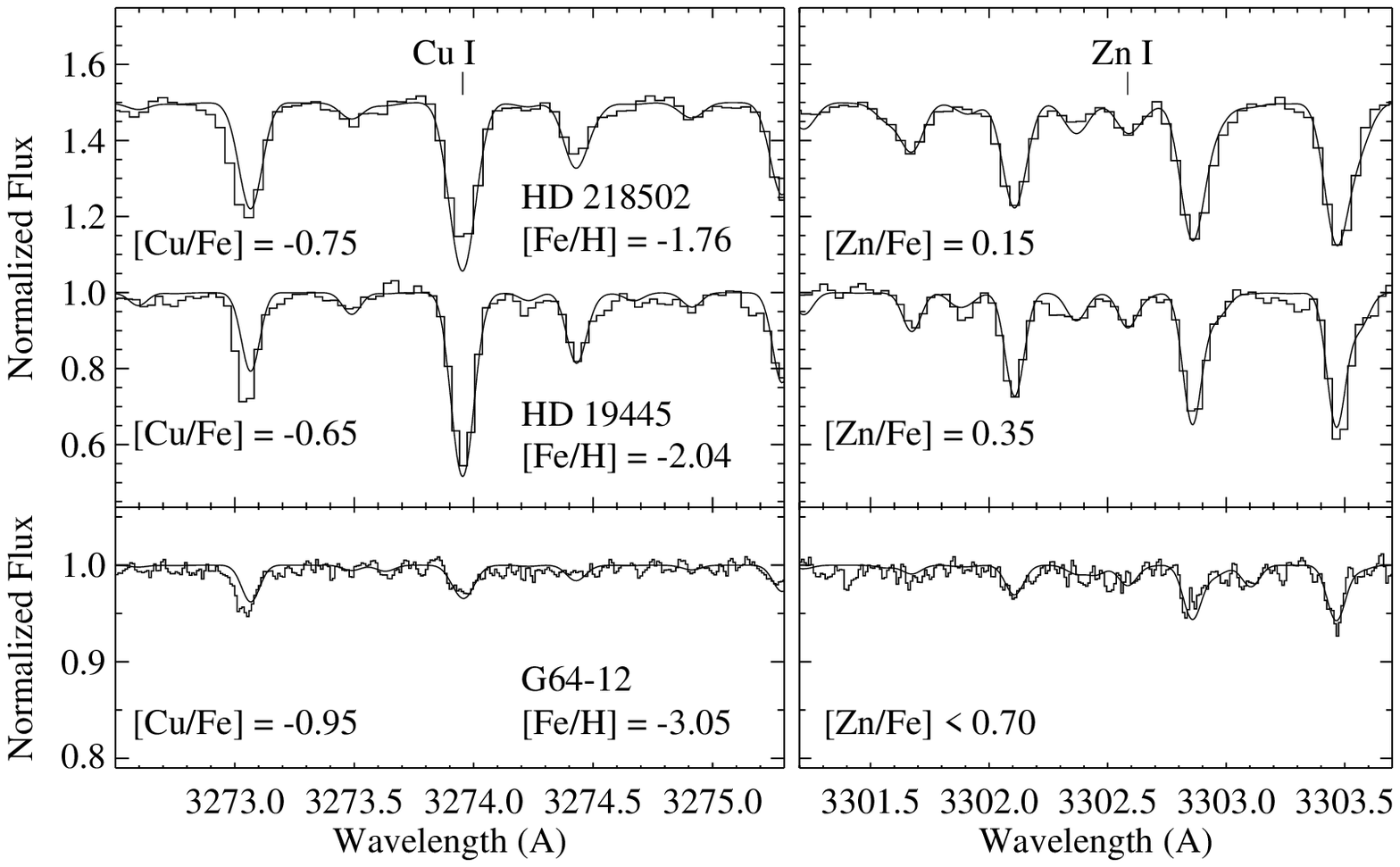}
\caption{{\bf Synthetic spectral fits to the copper and zinc
lines in the spectra of several metal-poor stars of our
sample.}} \label{fits} \end{figure*}

\begin{table*} \caption[]{Sensitivities of the copper and zinc
      abundances to variations in the atmospheric parameters of
      the stars \object{HD 94028} ($T_{\rm eff}$ $=$~6058~K,
      log~$g$ $=$~4.28, $V_{\rm t}$ $=$~1.0~km~s$^{-1}$),
      \object{HD 140283} ($T_{\rm eff}$ $=$~5723~K, log~$g$
      $=$~3.68, $V_{\rm t}$ $=$~1.0~km~s$^{-1}$) and
      \object{G170-47} ($T_{\rm eff}$ $=$~5154~K, log~$g$
      $=$~2.93, $V_{\rm t}$ $=$~1.25~km~s$^{-1}$), and to the
      overshooting:}

      \label{errors}
  $$	
     \begin{tabular}{l l l l l l l l l l l l l}
     	 \hline
	  &
	 \multicolumn{4}{l}{\object{HD 94028}} &
	 \multicolumn{4}{l}{\object{HD 140283}} &
	 \multicolumn{4}{l}{\object{G170-47}}
	 \\
	 &
	 $T_{\rm eff}$ &
	 log $g$ &
	 $V_{\rm t}$ &
	 Os &
	 $T_{\rm eff}$ &
	 log $g$ &
	 $V_{\rm t}$ &
	 Os &
	 $T_{\rm eff}$ &
	 log $g$ &
	 $V_{\rm t}$ &
	 Os 
	 \\
	 &
	 +100 &
	 +0.30 &
	 $-$0.3 &
	 off &
	 +100 &
	 +0.30 &
	 $-$0.3 &
	 off &
	 +100 &
	 +0.30 &
	 $-$0.3 &
	 off
	 \\
	 \hline
	 $[$Cu/H$]$ & +0.15 & $-$0.06 & +0.13 & $-$0.15 & +0.12 & 0.00  & +0.02 & $-$0.13 & +0.13 &  0.00 & +0.04 & $-$0.15\\
	 $[$Zn/H$]$ & +0.06 & +0.03 & +0.01 & $-$0.10 & +0.06 & +0.04 & 0.00  & $-$0.10 & +0.04 & +0.04 & 0.00  & $-$0.09\\
	 \hline \\
      \end{tabular}
  $$
\end{table*}

A few spectra also permitted the analysis of the other
\ion{Cu}{i} line at 3247.54~{\AA}. In the case of the very
metal-poor dwarfs \object{G64-12}, \object{G275-4}, \object{LP
815-43}, an agreement to within 0.15~dex was found between the
Cu abundances given by the two lines. Those in the spectrum of
the giant \object{BD $-$18$\degr$5550}, however, provided  a
difference of 0.25~dex, suggesting that the differential
analysis with respect to the \object{Sun} was less reliable for
this star (note that for the other giant, \object{HD 2796}, the
difference was still greater, reason why its Cu abundance was
not listed in Table~\ref{atmparam}). As for the zinc line, the
spectra of only six stars out of the thirteen with [Fe/H]
$<$~$-$2 permitted a clear detection. The spectra of the
remaining stars presented a very weak feature at 3302.58~\AA,
from which no reliable abundance could be derived; thus
upper-limits to the abundances were established.
Figure~\ref{fits} shows the fits for three metal-poor stars.\\

Since Kurucz's models atmospheres without overshooting 
reproduce better colour indices, Balmer profiles and Procyon
data than Kurucz's models with overshooting (Castelli et al.
\cite{caste97}), we checked the changes produced on the
abundances using the former models. For the stars \object{HD
94028}, \object{HD 140283} and \object{G147-70}
(Table~\ref{errors}), these models implied Cu and Zn abundances
systematically lower by about 0.15 and 0.10~dex, respectively,
differences small enough to preserve the main properties of our
results discussed in Sect.4.

As found by Cayrel et al. (\cite{cay03}), the abundances
derived from the spectra of metal-poor stars may be
overestimated if the continuum scattering is approximated as an
additional opacity source in the spectral synthesis code. The
overestimation is especially high for $\lambda$~$<$ 400~nm,
where continuum scattering becomes important relative to
continuous absorption. Since the spectral synthesis code MOOG
approximates scattering by absorption, we evaluated the
influence on the abundances by using another LTE spectral
synthesis code, TurboSpectrum (Alvarez \& Plez \cite{alv98}),
which takes proper account of the continuum scattering. For the
only two giants, \object{HD 2796} and \object{BD
$-$18$\degr$5550}, the Cu abundances derived using TurboSpectrum
were lower by more than 0.2~dex than those derived using MOOG,
while the Zn abundances were the same. As for the dwarfs
\object{G275-4} and \object{G64-12}, the abundances for both
elements were higher by less than +0.1~dex. We therefore
concluded that, for the dwarfs, the Cu and Zn abundances
obtained from the UV lines were not significantly affected by
this source of errors, while for the giants the Zn abundances
only. We thus removed the contribution of the giants from our
conclusion on the copper trend.

We compared our Cu and Zn abundances with those of Mishenina et
al. (\cite{mis02}) obtained from optical lines, and found a good
agreement: $\Delta$[Cu/H]~=
$\langle$[Cu/H]-[Cu/H]$_{M}$$\rangle$ =~$-$0.04~dex, with a
standard deviation $\sigma$ =~0.15~dex (16 stars; no giant data
to compare), and $\Delta$[Zn/H] =~0.03~dex, with $\sigma$
=~0.13~dex (16 stars), respectively. Converting our values with
our abundance sensitivities to their adopted stellar atmospheric
parameters provided a similar agreement: $\Delta$[Cu/H]
=~$-$0.09~dex, with $\sigma$ =~0.14~dex, and $\Delta$[Zn/H]
=~$-$0.02~dex, with $\sigma$ =~0.13~dex, respectively. The
slight abundance decrease was due principally to their lower
effective temperatures and microturbulence. The dwarf \object{HD
19445} (see Fig.~\ref{fits}) was the most metal-poor star for Cu
abundance comparison with Mishenina et al. (\cite{mis02}), with
[Fe/H] =~$-$2.04. We found [Cu/H] =~$-$2.69, a value  0.37~dex
lower than that of Mishenina et al. (\cite{mis02}), and lower by
$\sim$0.1~dex than the upper limit estimate of Sneden et al.\
(\cite{sne91}) (for their adopted atmospheric parameters). We
also compared our results with the abundances obtained from
optical lines by Sneden et al (\cite{sne91}), and we found a
rough agreement for Cu, $\Delta$[Cu/H] =~0.13~dex, with $\sigma$
=~0.36~dex (5 stars only\footnote{The stars were less metal-poor
than \object{HD 19445} and common to the sample of Mishenina et
al. (\cite{mis02}).}; no giant data to compare), and for Zn,
$\Delta$[Zn/H] =~0.13~dex, with $\sigma$ =~0.13~dex (8 stars),
respectively. Converting our values with our abundance
sensitivities to their adopted stellar atmospheric parameters
also revealed  a rough agreement for Cu, $\Delta$[Cu/H]
=~0.17~dex, with $\sigma$ =~0.24~dex, while a good agreement for
Zn, $\Delta$[Zn/H] =~0.04~dex, with $\sigma$ =~0.14~dex,
respectively.\\

\section{Trends of [Cu/Fe] and [Zn/Fe]}

\subsection{Results}

\begin{figure*}
\centering
\includegraphics[width=17cm]{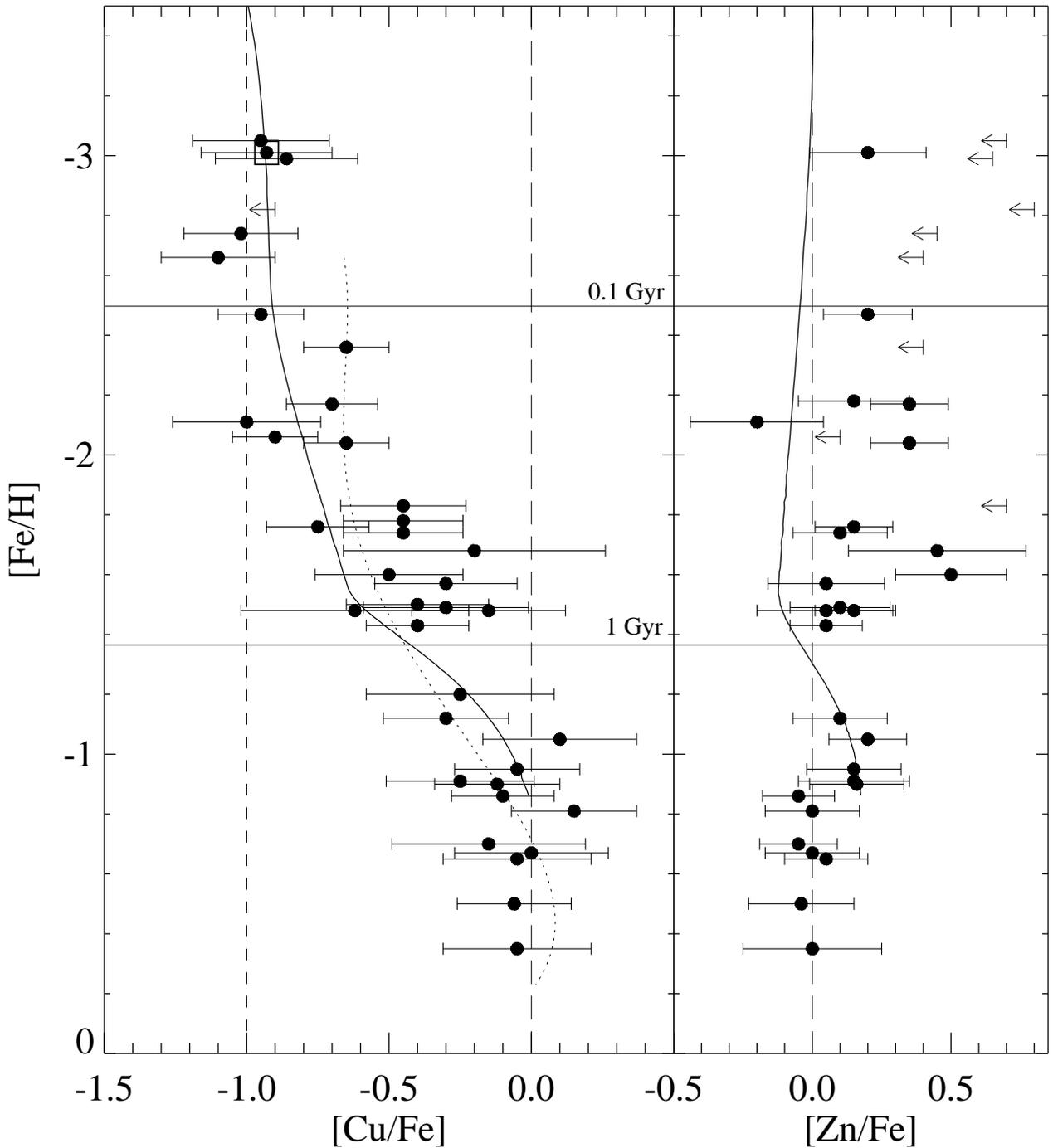}
\caption{[Cu/Fe] and [Zn/Fe] ratios vs. [Fe/H]. The upper
limits of zinc abundances are represented  by the {\it arrows},
the copper abundance in the giant \object{BD $-$18$\degr$5550}
by the {\it square} and the fit to the copper data obtained by
Mishenina et al. by the {\it dotted line}. The {\it solid lines}
represent the predictions of Prantzos models and the {\it
horizontal lines} represent the ellapsed time since the
beginning of the halo formation.} \label{trendCuZn}
\end{figure*}

Our derived [Cu/Fe] and [Zn/Fe] ratios are shown in
Fig.~\ref{trendCuZn}. The [Cu/Fe] ratio is almost solar down to
[Fe/H] $\sim$~$-$1 and then decreases, until reaching a plateau
$\langle$[Cu/Fe]$\rangle$ =~$-$0.98  at [Fe/H] $<$~$-$2.5 (the
ratio is averaged over the five stars that are not giants). This
slanted s-shape trend in dwarfs is similar to that outlined by
Mishenina et al. (\cite{mis02}) (see Fig.~\ref{trendCuZn}).
However, the [Cu/Fe] plateau is $\sim$~0.3~dex lower than the
plateau obtained in the range $-$2.7 $<$~[Fe/H] $<$~$-$1.7 in
their halo giants and the globular clusters studied by Simmerer
et al. (\cite{sim03}). It could be noticed that underabundances
near [Cu/Fe] =~$-$1 are also found in Mishenina et al.
(\cite{mis02}) (the two dwarfs \object{BD +41$\degr$3931} and
\object{HD 140283}\footnote{In this reference, no [Cu/Fe]  value
is given for \object{HD 140283}. We refer to the value derived
from the point at the corresponding metallicity in Fig. 9.},
Fig. 9, with [Cu/Fe] =~$-$0.97 and $\sim$~$-$0.95, respectively)
and in Sneden et al. (\cite{sne91}) (the giants \object{HD 2665}
and \object{HD 122563}, Fig. 7, with [Cu/Fe] =~$-$0.87 and
$-$0.93). In addition, Sneden et al. (\cite{sne91}) provides an
upper limit [Cu/H] $<$~$-$3.93 for the giant \object{BD
$-$18$\degr$5550}, and we find [Cu/H] =~$-$3.94 ([Cu/Fe]
=~$-$0.93), a value that would be even lower if we were to
take into account a better treatment of the UV spectral
synthesis (see Subsection~\ref{Cu and Zn Abundances}). Thus,
underabundances near [Cu/Fe] =~$-$1 in very metal-poor dwarfs
and giants should not be considered infrequent.\\

\begin{figure}
\resizebox{\hsize}{!}{\includegraphics{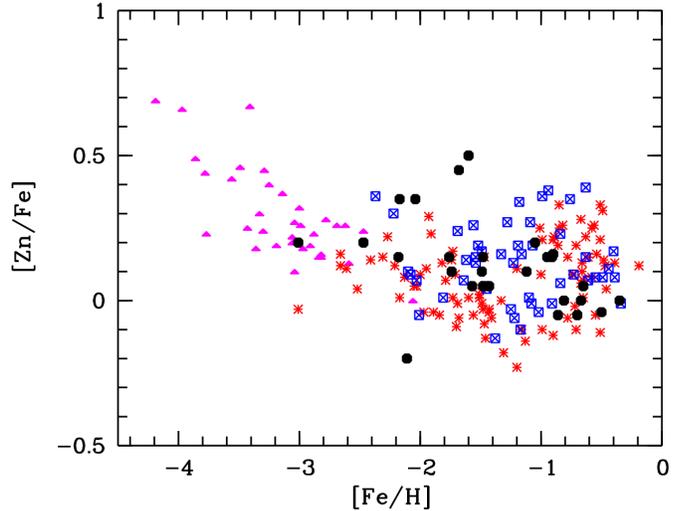}}
\caption{[Zn/Fe] ratios vs. [Fe/H] from our measurements ({\it
      filled dots}), from Cayrel et al. ({\it filled
      triangles}), from Mishenina et al. ({\it asterisks}) and
      Gratton et al. ({\it crossed squares}). The [Zn/Fe] values
      of Gratton et al. are decreased by 0.09~dex to be
      normalized to meteoritic rather to solar abundances.}
\label{trendZn_alldata}
\end{figure}

Considering zinc, the [Zn/Fe] ratio is essentially solar down to
[Fe/H]$\sim -2.0$, with a possible excess of
$\langle$[Zn/Fe]$\rangle$ =~+0.18 at lower metallicities. It is
interesting to compare our results with existing high quality
surveys which include observations of Zn: those of Mishenina et
al. (\cite{mis02}) and Gratton et al. (\cite{gra03}) in a
metallicity range comparable to that of this study, and that of
Cayrel et al. (\cite{cay03}), which is complementary and covers
considerably lower metallicities. The comparison of our data
with those of these three studies is shown in
Fig.~\ref{trendZn_alldata}. The ensemble of all the data gives
for the first time a clear picture of the evolution of the Zn/Fe
ratio with metallicity. A solar ratio of Zn/Fe is supported by
the data of both Mishenina et al. (\cite{mis02}) and Gratton et
al. (\cite{gra03}) down to a metallicity of about [Fe/H] $\sim
-2.5$. The rise, which is hinted at by the lower metallicity
tail of our sample, is consistent with the robust rise from the
Cayrel et al. (\cite{cay03}) sample. The regression line for
[Zn/Fe] provided in Table 7 of Cayrel et al. (\cite{cay03})
intercepts the solar ratio for [Fe/H]$\sim -2.1$. It is worth
noting that our data are consistent with the data of both
Mishenina et al. (\cite{mis02}) and Gratton et al.
(\cite{gra03}) over the common metallicity range, and that our
Zn abundances are derived from the \ion{Zn}{i} 3302.58~\AA line,
while the other groups use optical lines. Two stars,
\object{HD 166913} and \object{HD 132475}, however, seem to defy
agreement, by presenting [Zn/Fe] greater than 0.45. In the case
of \object{HD 166913}, the apparent high ratio [Zn/Fe] =~0.45
stems not from the Zn abundance itself, which is 0.11~dex above
that derived by Gratton et al. (\cite{gra03}) (once  the
different atmospheric parameters are taken into account, using
the abundance sensitivities given in Table~\ref{errors} for
\object{HD 94028}), but rather from our much lower Fe abundance,
adopted from Th\'evenin et al. (\cite{the98}) and 0.23~dex lower
than Gratton et al.'s (\cite{gra03}). For the latter comparison,
we use the abundance sensitivities from Gratton et al.
(\cite{gra03}; Table 9), since none are given in Th\'evenin et
al. (\cite{the98}). In the case of \object{HD 132475}, the high
ratio [Zn/Fe] =~0.50 stems from its reddening, $E(b-y)$ = 0.046
(Schuster \& Nissen \cite{sch89}), because this reddening
implies an appreciably higher effective temperature and thus a
lower ratio [Zn/Fe]  (if referred to Table 9, Gratton et al.
\cite{gra03}).

\begin{figure}
\resizebox{\hsize}{!}{\includegraphics{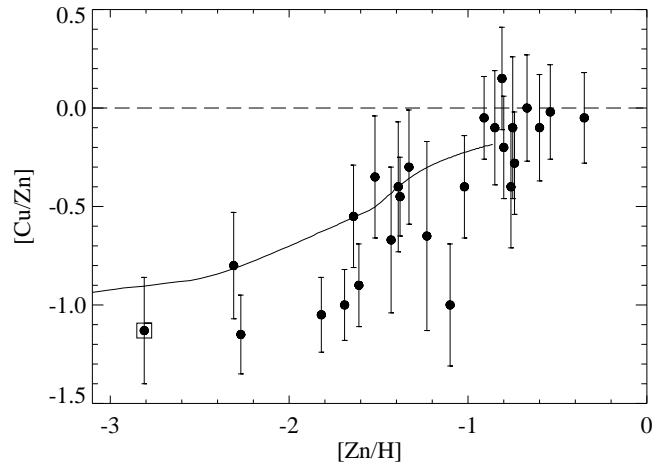}}
\caption{[Cu/Zn] ratios vs. [Zn/H]. The {\it solid line}
represents the predictions of Prantzos models.}
\label{trendCu_vs_Zn} \end{figure}

\subsection{Cu and Zn nucleosynthesis}

In spite of considerable theoretical effort, the production
sites of Cu and Zn are not clear (e.g. the discussion in
Mishenina et al. \cite{mis02}, and references therein).\\

- From the theoretical point of view both elements may be
synthesized in massive stars. Both may be formed through
n-capture during He-burning in hydrostatic equilibrium (Woosley
\& Weaver \cite{woo95}). In addition, Cu may also be formed in
explosive Ne-burning (Woosley \& Weaver \cite{woo95}), while Zn
may be formed in explosive Si-burning and consequent
$\alpha$-rich freeze-out (Arnett et al. \cite{arn71}). A minor
contribution to both elements may come from $s$-process in
intermediate mass stars (Gallino et al. \cite{gal98}). Also,
Type Ia SNe may contribute their fraction of Cu and Zn (Iwamoto
et al. \cite{iwa99}). Finally, there are also suggestions that
there might be contributions from the $r$-process (Woosley \&
Weaver \cite{woo95}; Umeda \& Nomoto \cite{ume02}).\\

The relative importance of all these potentially relevant
processes remains to be evaluated.\\

- From the observational point of view some of these
nucleosynthetic channels may be either ruled out or at least
relegated to a very minor role.\\

The very different behavior of Cu and Zn implies that n-capture
during He-burning cannot be the main mode of production for
both. Concerning the production of Zn, there are reasons to
believe that neither the $s$- nor the $r$-process can be very
relevant (Cayrel et al. \cite{cay03}). If the $s$-process were
important, the [Zn/Fe] ratio ought to decrease with decreasing
metallicity, as is the case for typical $s$-process products
such as Y, Sr or Ba. From Fig.~\ref{trendZn_alldata}, it is
instead clear that this ratio is initially constant and then
{\em increases} at lower  metallicities. This observational
argument supports the conclusions drawn from the theoretical 
computations of Gallino et al. (\cite{gal98}). Concerning the
relevance of the $r$-process, we may note that in the extremely
metal-poor star \object{CS 31082-001} ([Fe/H] =~$-$2.90), in
which all the $r$-process elements are greatly enhanced, [Zn/Fe]
=~+0.18 (Hill et al. \cite{hil02}), remarkably similar to the
behaviour of other stars of comparable metallicity.\\ 

 We may  also extend the above argument to Cu; in fact, no Cu is
detected in CS 31082-001, although no upper limit is published.
A greatly enhanced Cu would surely be detected in the high
quality data of Hill et al. (\cite{hil02}). Therefore we may
conclude that strong Cu production in the $r-$process is
unlikely.\\

In this study we compare our results with new chemical evolution
predictions from Prantzos (private communication, \cite{pra03}),
which are following those from Goswami \& Prantzos
(\cite{gos00}). The models are based on the stellar initial mass
function (IMF) from Kroupa et al. (\cite{kro93}) and assume an
evolution of the halo for a gas outflow rate which is equal to 8
times the star formation rate. We find that the Prantzos model
(\cite{pra03}) most reasonably agrees with the observations -
especially for Cu, except for some points (see
Fig.~\ref{trendCuZn}). Model and observations agree on the
existence of a plateau for Cu at very low metallicities ([Fe/H]
$<$~$-$2.5). This supports the adopted SNII yields
(Woosley \& Weaver \cite{woo95}) averaged over the stellar
initial mass function from Kroupa et al. (\cite{kro93}).
Nevertheless, we stress that since higher ratios than those
predicted are also observed at low metallicity (Sneden et al.
\cite{sne91}; Mishenina et al. \cite{mis02}), we can only
conclude on a possibly lower limit for the relative production
of copper. As for Zn, we observe local agreements and a
systematic disagreement appearing at [Fe/H] $<$~$-$2,
indicating, to a first approximation, that some Zn sources are
missing.\\

\section{Conclusions}

In this study, we present Cu and Zn abundances for a sample of
38 stars, principally dwarfs, with metallicity ranging from
solar to $-$3. The abundances were obtained using for the first
time the intense UV lines \ion{Cu}{i} 3273.95~{\AA} and
\ion{Zn}{i} 3302.58~{\AA}, and were generally in agreement with
the results from previous studies. We showed that the Cu
abundance indicator is adequate for the study of abundances in
very metal-poor stars, permitting to detect [Cu/Fe] $\sim$~$-$1
in dwarfs at [Fe/H] =~$-$3. This research can be extended to
dwarf stars of even lower metallicities and with high resolution
spectra. Contrary to the slanted s-shape trend found for
[Cu/Fe], the [Zn/Fe] trend is approximately solar in most stars
with [Fe/H] $>$~$-$2; this is a common result, which supports
the use of Zn as a metallicity tracer in damped Ly$\alpha$
systems. However, at [Fe/H] $<$~$-$2, Zn appears to be slightly
overabundant ($\langle$[Zn/Fe]$\rangle$ =~$+$0.18) and it is in
fact at these metallicities that the rise also becomes
noticeable in the sample of stars studied by Cayrel et al.
(\cite{cay03}).\\

The trends found can be reasonably reproduced by the halo
chemical evolution predictions from Prantzos (\cite{pra03},
private communication). The ratio [Cu/Fe] $\sim$~$-$1 for $-$3
$<$~[Fe/H] $<$~$-$2.5 is predicted in Prantzos model for SNII
yields from Woosley \& Weaver (\cite{woo95}). It probably
represents, at this evolutionary stage, a lower limit of Cu
production, since greater ratios are also observed in previous
studies. The [Zn/Fe] trend is in general in good agreement with
that predicted, except in the low-metallicity tail, where the
predicted ratios should be systematically increased.\\

\begin{acknowledgements}

We would like to thank Peter A. Bergbusch and Don A. VandenBerg
for providing the software producing the isochrones, and Fran{\c
c}ois Th\'evenin for  information about the Catalogue III/193.
We  would also like to thank Robert L. Kurucz for  information
about the intense copper lines and their hyperfine structure,
Nicolas Prantzos for providing the unpublished trends of copper
and zinc, and Bertrand Plez for the TurboSpectrum code. This
research has made use of the SIMBAD database, operated at CDS,
Strasbourg, France, and also of data results from the Two Micron
All Sky Survey, which is a joint project of the University of
Massachusetts and the Infrared Processing and Analysis Center,
funded by the National Aeronautics and Space Administration and
the National Science Foundation.

\end{acknowledgements}

\end{document}